%% file: column.tex
\documentstyle[amssymb,epsfig,twoside]{europhys}
\input euromacr

\newcommand{\bA}{{\bf A}}
\newcommand{\bB}{{\bf B}}
\newcommand{\bH}{{\bf H}}

\newcommand{\br}{{\bf r}}

\newcommand{\bm}{{\bf m}}

\newcommand{\f}{\frac}

\newcommand{\Eq}[1]{Eq.~(\ref{#1})}

\newcommand{\rmp}{\Review{Rev.\ Mod.\ Phys.}}
\newcommand{\prl}{\Review{Phys.\ Rev.\ Lett.}}
\newcommand{\prb}{\Review{Phys.\ Rev.\ B}}

\begin{document}
\euro{47}{4}{494-500}{1999}
\Date{15 August 1999}
\shorttitle{J.\ LIDMAR and M.\ WALLIN: CRITICAL PROPERTIES OF BOSE-GLASS SUPERCONDUCTORS}

\title{Critical properties of Bose-glass superconductors}

\author{Jack~Lidmar and Mats~Wallin} 

\institute{Department of Theoretical Physics, Royal Institute of
Technology\\ SE-100 44 Stockholm, Sweden}

\rec{2 June 1999}{28 June 1999}

\pacs{
\Pacs{74}{60.Ge}{Flux pinning}
\Pacs{75}{40.Mg}{Num.\ Simulations}
\Pacs{05}{70.Fh}{Phase Trans}
}

\maketitle

\begin{abstract}
We study vortex lines in high-temperature superconductors with
columnar defects produced by heavy ion irradiation.  We reconsider
scaling theory for the Bose glass transition with tilted magnetic
fields, and propose, e.g., a new scaling form for the shape of the
Bose glass phase boundary, which is relevant for experiments.  We also
consider Monte Carlo simulations for a vortex model with a screened
interaction.  Critical exponents are determined from scaling analysis
of Monte Carlo data for current-voltage characteristics and other
quantities.  The dynamic critical exponent is found to be $z = 4.6 \pm
0.3$.
\end{abstract}

A rich variety of new vortex phase transitions has been established in
high-temperature superconductors~\cite{Blatter}.  Bose glass physics
was originally suggested, by Nelson and Vinokur~\cite{nelson-vinokur},
to apply for vortex phase transitions in systems with artificially
introduced columnar defects, produced as permanent damage tracks from
heavy ion irradiation of the sample.  Columnar defects act as optimal
pinning centers, leading to a considerable increase of critical
currents and fields, as compared to the unirradiated
sample~\cite{civale,nelson-vinokur,lee-stroud-girvin}.  Bose glass
theory makes a set of distinct predictions, e.g., for the dynamic
scaling of vortex transport properties, and the response to tilted
magnetic fields~\cite{nelson-vinokur,wallin-girvin,radzihovsky}.  Such
behavior has been observed in various
experiments~\cite{budhani,jiang,reed,seow,phuoc,grigera}.

Vortex lines in 3D superconductors with columnar defects can be mapped
to imaginary time world lines of bosons in 2+1 dimensions on a
disordered substrate~\cite{nelson-vinokur}.  The superconducting glass
phase for the vortex lines corresponds to the insulating Bose glass
(BG) phase for the bosons.  The dissipative vortex line liquid phase
corresponds to the superconducting phase for bosons.  This mapping
gives useful information about the equilibrium properties in the
columnar defect problem \cite{dirtybosons}.  However, dynamical
properties do not follow from the mapping.  A tilted magnetic field
$H_\perp$ with respect to the columns enters like an imaginary vector
potential for the bosons \cite{nelson-vinokur}, leading to a
localization problem in non-Hermitian quantum mechanics, which has
received considerable attention recently \cite{hatano-nelson,efetov}.
The vortices want to stay localized on the columns, which results in a
transverse Meissner effect with a divergent tilt modulus $c_{44}$.
The BG phase boundary in the $(T, H_{\perp})$-plane has a sharp cusp
at $H_{\perp}=0$, which distinguishes the Bose glass from the
isotropic point-defect vortex glass which has a smooth phase boundary
\cite{nelson-vinokur}.  Such a feature in the transition line is seen
in experiments~\cite{jiang,reed,grigera}, and provides support for the
Bose glass theory.  Other types of correlated disorder, like splayed
columnar defects~\cite{splay}, and planar
defects~\cite{nelson-vinokur}, also strongly influence the properties
of the glass phase.

In this paper we reconsider scaling theory for the Bose glass
transition for the case of tilted magnetic fields, which leads to
certain important modifications of predictions in earlier
work~\cite{nelson-vinokur}, e.g., for the form of the BG phase
boundary.  We further present Monte Carlo (MC) simulations for various
thermodynamic and dynamic quantities, which verifies the scaling
predictions derived below, and give an estimate of the dynamical
critical exponent $z$.

We first discuss scaling theory.  A detailed scaling theory has been
developed for the superconducting phase transition by Fisher et al.\
\cite{ffh91}, and generalized to anisotropic Bose glass scaling in
Refs.\ \cite{nelson-vinokur,wallin-girvin,radzihovsky}.  In the
present problem it is crucial to correctly distinguish between the
scaling of the $\bB$ and $\bH$ fields.  These relations have sometimes
been mixed up in the literature.
On approaching the transition the correlation lengths in the
directions perpendicular and parallel to the columnar defects are
assumed to diverge as $\xi \equiv \xi_{\perp} \sim |T-T_{\rm
BG}|^{-\nu}$ and $\xi_{\parallel} \sim \xi^\zeta$, respectively, where
$\zeta$ is an anisotropy exponent.  Due to screening of the
interaction and the correlated disorder, the correlation volume is
anisotropic with $\zeta = 2$~\cite{dirtybosons}.  In case of
unscreened long-range interactions we expect instead $\zeta \approx
1$.  The correlation time diverges as $\tau \sim \xi^z$ where $z$ is
the dynamical critical exponent.  The vector potential enters in the
combination $\nabla - (2\pi i/\Phi_0)\bA$, and therefore scales as
$A_\perp \sim \xi^{-1}$, $A_\parallel \sim \xi_\parallel^{-1}$.  From
hyperscaling the free energy density scales as $f \sim
\xi^{1-d-\zeta}$, where
$d$ is the dimension,
and therefore the current density
scales as $J_{\perp}={\partial f \over \partial A_{\perp}} \sim
\xi^{2-d-\zeta}, J_{\parallel}={\partial f \over \partial
A_{\parallel}} \sim \xi^{1-d}$, and the electric field as $E_{\perp} =
- \f{1}{c}\f{\partial A_{\perp}}{\partial t} \sim \xi^{-(1+z)},
E_{\parallel} = - \f{1}{c}\f{\partial A_{\parallel}}{\partial t} \sim
\xi^{-(\zeta+z)}$.  The scaling of the magnetic field is obtained from
$B=-4\pi \f{\partial f}{\partial H}$ (or from $\nabla \times
\bH=\f{4\pi}{c} \mathbf{J})$, which gives $H_{\perp} \sim \xi^{2-d},
H_{\parallel} \sim \xi^{3-d-\zeta}$, and the flux density scales as
$B_{\perp} = (\nabla \times \bA)_{\perp} \sim \xi^{-1-\zeta},
B_{\parallel} = (\nabla \times \bA)_{\parallel} \sim \xi^{-2}$.  The
appropriate scaling combinations involving the magnetic field are
therefore $H_{\perp} \xi^{d-2}$ and $H_{\parallel} \xi^{d-3+\zeta}$,
which differs from those found, e.g., in Ref.~\cite{nelson-vinokur}.
This has several experimentally relevant consequences as we will
discuss below.  The linear resistivity scales as
\begin{eqnarray}
    \rho_{\perp} &=& \xi^{d+\zeta-3-z} \tilde\rho_\pm^{\perp}(H_{\perp} \xi^{d-2})
    \label{rho_lin-perp}
\\
    \rho_{\parallel} &=& \xi^{d-\zeta-1-z} \tilde\rho_\pm^{\parallel}(H_{\perp} \xi^{d-2}),
    \label{rho_lin-para}
\end{eqnarray}
and the I-V characteristic as
\begin{eqnarray}
    E_{\perp} &=& \xi^{-(1+z)} \tilde{E}_{\pm}^{\perp}
    (J_\perp\xi^{d+\zeta-2},H_{\perp} \xi^{d-2})
\label{I-V-perp}
\\ 
    E_\parallel &=& \xi^{-(\zeta+z)} \tilde{E}_{\pm}^{\parallel}
    (J_\parallel\xi^{d-1},H_{\perp} \xi^{d-2}),
\label{I-V-para}
\end{eqnarray}
where $\tilde{\rho}$ and $\tilde{E}$ are scaling functions.
For the magnetic permeability we obtain
\begin{eqnarray}
   \mu_\perp &=& \xi^{d-3-\zeta} \tilde{\mu}_{\perp}
   (H_{\perp}\xi^{d-2})
\label{mu-perp}
\\ 
   \mu_\parallel &=& \xi^{d-5+\zeta} \tilde{\mu}_{\parallel}
   (H_{\perp}\xi^{d-2}).
\label{mu-para}
\end{eqnarray} 
As an independent check of the correctness of this scaling law we can
again use the mapping to dirty bosons in (2+1)D.  The permeability
$\mu_\perp$ corresponds to the superfluid density $\rho_s$ for the
bosons,
whose scaling was derived in Ref.~\cite{dirtybosons} with the result
$\rho_s \sim \xi^{3-d-\zeta}$.
This scaling law coincides with Eq.\ (\ref{mu-perp}) for $d=3$
\cite{foot}.  Equation (\ref{mu-para}) corresponds to the
compressibility of the bosons $\kappa\sim \xi^{\zeta-d+1}$,
again with agreement for $d=3$.
  To obtain
the scaling of the BG phase boundary for small tilt, Eq.\
(\ref{mu-perp}) gives $\mu_\perp = t^{-\nu(d-3-\zeta)}
\tilde{\mu}_{\perp} (H_{\perp}t^{-\nu(d-2)})$, where $t=|T-T_{\rm
BG}|$, and following Ref.\ \cite{nelson-vinokur} we assume that the
Bose glass phase persists up to a finite tilt, so
$\tilde{\mu}_{\perp}(x)$ has a singularity at a finite value $x=x_c$.
At this point the critical tilt field vanishes when $T \to T_{\rm BG}$
as
\begin{equation}
   T_{\rm BG}-T \sim |H_{\perp}|^{1/\nu},
\label{line}
\end{equation}
for $d=3$.  Equations (\ref{mu-perp}-\ref{line}) differ from the
corresponding ones in Ref.\ \cite{nelson-vinokur}, and in particular
our Eq.\ (\ref{line}) does not include a factor 3 in the exponent.
The shape of the phase boundary still has a cusp at $H_\perp=0$, but
not as sharp as the earlier predictions suggested.
Another consequence of experimental interest is the scaling relation
$\rho_\perp \sim H_\perp^{z-\zeta}$ for the resistivity at $T=T_{\rm
BG}$.

We now turn to our Monte Carlo simulations.  We concentrate on low
magnetic fields, $B \lesssim \Phi_0/\lambda^2$, where the vortex
interaction is effectively short ranged, and also restrict our study
to fields below the matching field, where the density of columnar
defects is greater than the vortex density.  Vortex dynamics has been
studied previously in simulations of the nonlinear current-voltage
(I-V) characteristics~\cite{wallin-girvin} for finite applied
supercurrents.  In this paper we present extensive simulation results
for the linear resistance, and compare with dynamical scaling of the
I-V characteristics.  We choose to study the dirty Boson action with
an onsite repulsion \cite{dirtybosons}, which gives a coarse-grained
representation of the physical system,
\begin{equation}
   H = \sum_{\bf r} \left\{ {1 \over 2} \bm_{\br}^2 
   - g({{\bf r}_{\perp}}) m_{\br z} \right\},
\label{screen}
\end{equation}
where the integer variable $\bm = (m_x,m_y,m_z)$ is the vorticity
vector on the links of a simple cubic lattice of size $\Omega=L\times
L\times L_{\parallel}$ with lattice constant $a=1$.  The partition
function is $Z={\rm Tr}\, e^{-\beta H}$, where the trace is the sum
over all vortex line configurations with no open ends, thus satisfying
the constraint $\nabla\cdot\bm=0$, and $\beta=1/T$ is the inverse
temperature.  Periodic boundary conditions are used in all directions
to eliminate surface effects.  The columnar defects are aligned in the
$z$-direction, and are modeled as a uniformly distributed random site
energy $g({\bf r}_{\perp})$, which is constant in the $z$-direction.
The applied magnetic field is included as a finite fixed net density
of vortex lines.  We also consider tilted magnetic fields by allowing
global vortex line fluctuations in the direction perpendicular to the
columns, with a bias term $-\sum_\br H_{x} m_{\br x}$ added to the
Hamiltonian.  Although not realistic in detail, this model should fall
into the universality class of strongly screened vortices with
columnar defects and therefore give the critical exponents correctly
for low enough magnetic fields.

Our Monte Carlo method for vortex line models has been described
elsewhere and is only outlined here~\cite{lidmar98}.  The starting
configuration is taken as an assembly of straight vortex lines
penetrating the system in the $z$-direction.  The MC moves
are attempts to add closed vortex loops around randomly selected
elementary plaquettes of the lattice.  The trial
moves are accepted with probability $1/(1+e^{\beta\Delta H})$, where
$\Delta H$ is the total change in energy.

I-V characteristics can be calculated by the following
method~\cite{wallin-girvin,lidmar98}.  Each time a loop is formed it
generates a voltage pulse $\Delta Q = \pm 1$ perpendicular to its
plane, the sign depending on the orientation of the loop.  This leads
to a net electric field $E(t) = \frac{h}{2e} J^V(t)$, (in the
following we set $h/(2e)=1$) where the vortex current density at MC
time $t$ is given by $J^V(t) = \frac{\Delta Q}{\Delta t}$, and $\Delta
t = 1$ for one full sweep through the system, where, on average, an
attempt is made to create or destroy one vortex loop on every
plaquette of the lattice.  The nonlinear I-V characteristic can be
modeled as the electric field $E$, due to vortex current response in
the presence of a uniform Lorentz force on the vortex lines,
proportional to the applied current density $J$.  The linear
resistance can be calculated from the equilibrium voltage fluctuations
via the Kubo formula~\cite{young94}, $ R = \frac{1}{2T}
\sum_{t=-\infty}^{\infty} \Delta t \, [ \langle V(t)V(0) \rangle ].  $
Here $\langle \cdots \rangle$ denotes thermal average and $[\cdots]$
disorder average.

To study the response to tilt, we consider an ensemble with
fluctuating winding number.  We achieve this by including global MC
moves where vortex lines are inserted across the whole system in the
direction perpendicular to the columns.  The uniform part of the flux
density is proportional to the winding number, $B_\nu = {W_\nu \over
\Omega} = {1 \over \Omega}\sum_\br m_{\br\nu}$.  In a tilted magnetic
field $[\left<B_\perp\right>]$ is non-zero above the BG temperature
and zero below.  The magnetic permeability in the direction
perpendicular to the columns is given by
\begin{equation} \label{perm}
   \mu_{\nu} = \frac{\partial \left[\left< B_{\nu} \right>\right]}
   {\partial H_{\nu}} = {\Omega \over T} \left[ \left< B_{\nu}^2
   \right> - \left< B_{\nu} \right>^2 \right],
\end{equation}
and is related to the tilt modulus by $c_{44} \sim 1/\mu_{\perp}$.  In
the analogy to dirty bosons, $\mu_{\perp}$ corresponds to the
renormalized superfluid density $\rho_s$ of the bosons.

The details of our MC calculations are as follows.  We use a net
density of vortex lines corresponding to filling of $f=1/2$ flux
quanta per plaquette.  We use a disorder potential $g$ uniformly
distributed in $[0, 1]$. The model, Eq.\ (\ref{screen}), has been
extensively studied in the context of dirty bosons, and has a glass
transition at $T_{\rm BG} \approx 0.248$ \cite{dirtybosons} (in case
of no tilted magnetic field). In finite systems the scaling functions
obtain extra arguments $\xi/L$ and $\xi_{\parallel}/L_{\parallel}$.
To enable finite size scaling of MC data we use system sizes $L_z = c
L^\zeta$, making the ratio $L_{\parallel}/L^\zeta$ a constant, which
eliminates one argument from the scaling functions.  We use $\zeta=2$,
$c=1$, and lattice sizes $L=4, \dots, 16$.  For each realization of
the columnar disorder, up to $2 \cdot 10^4$ MC sweeps were discarded
for warmup, and measurements taken during up to $2 \cdot 10^6$ sweeps.
Up to $10^3$ disorders were used to obtain small statistical errors in
the disorder averages.

\begin{figure}[t]
\centerline{
\epsfxsize 0.5\columnwidth \epsfbox{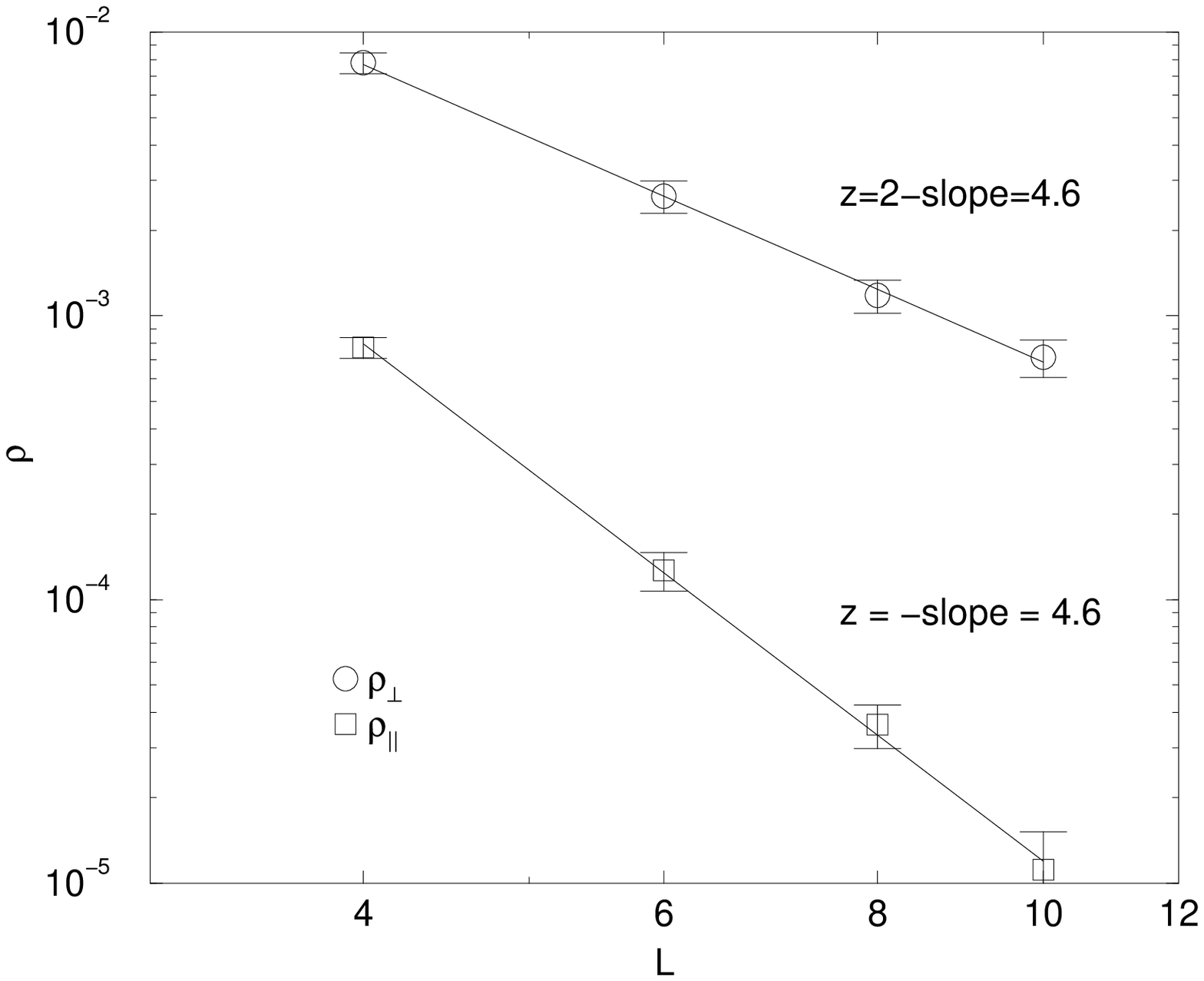}
\epsfxsize 0.5\columnwidth\epsfbox{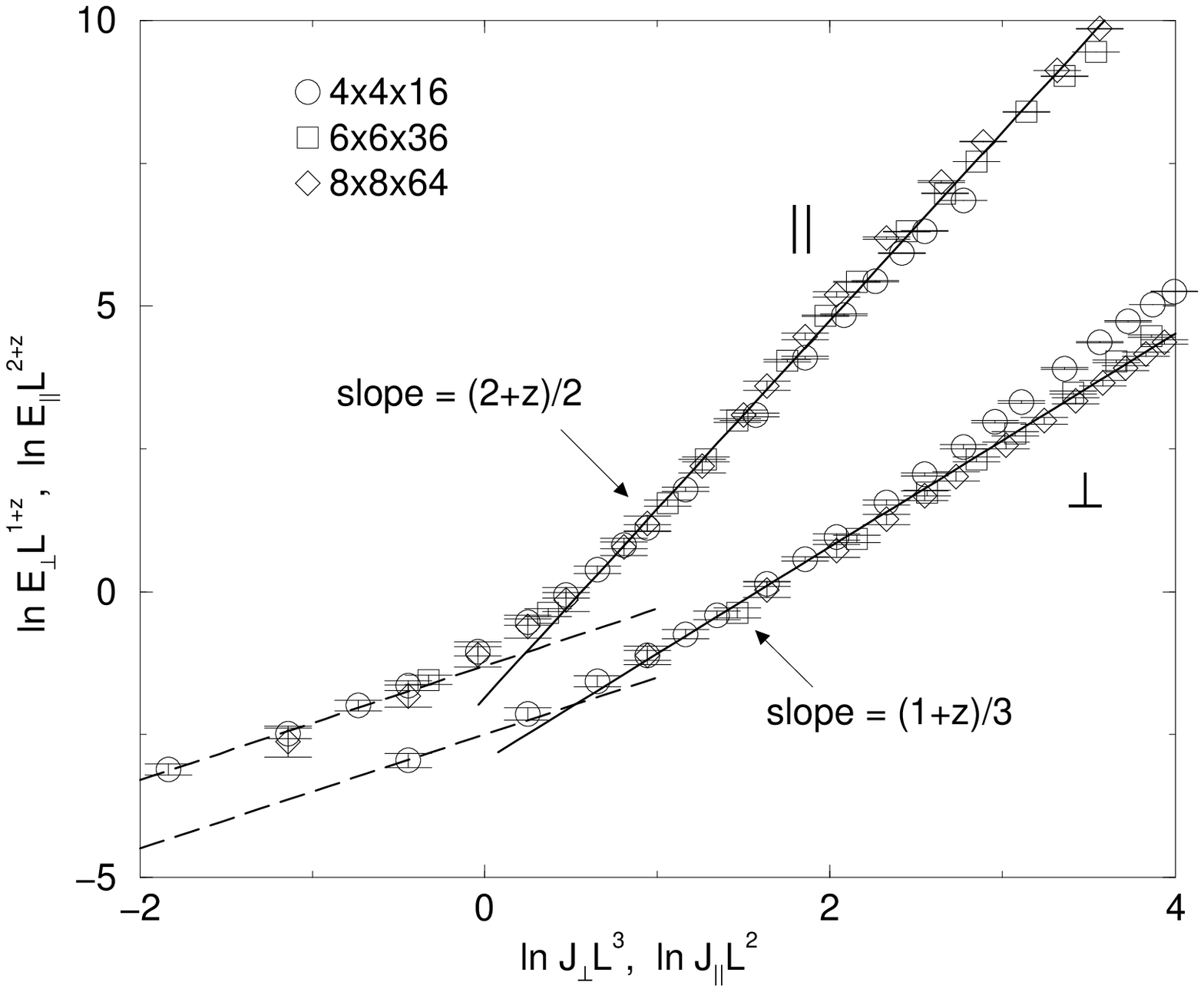}}
\caption{MC results for the linear resistivity vs.\ system size $L$ at
the Bose glass transition in the directions parallel and perpendicular
to the columnar defects.  From the slope of the curves the dynamical
critical exponent is estimated to be $z\approx 4.6$ for both
directions.  }
\label{rho}
\end{figure}

\begin{figure}
\caption{Nonlinear current-voltage characteristics at the Bose glass
critical point.  Dashed lines have slope
one (Ohmic response).  Solid lines have slope corresponding to the
value $z=4.6$ from the linear resistivity in Fig.\ 1.  }
\label{IV}
\end{figure}

We now turn to our MC results, starting with the magnetic field
applied along the columns.  Figure \ref{rho} shows a log-log plot of
the linear resistivities $\rho_{\perp}$ in the direction perpendicular
to the columnar defects, and $\rho_{\parallel}$ parallel to the
columns, vs.\ system size $L$ at $T=T_{\rm BG}$.  Power law fits to
the data (solid lines) allow $z$ to be determined according to
Eqs.~(\ref{rho_lin-perp},\ref{rho_lin-para}), which gives $z = 4.6 \pm
0.3$ in both directions.

Figure \ref{IV} shows a finite size scaling data collapse according to
Eqs.\ (\ref{I-V-perp},\ref{I-V-para}) of the nonlinear current-voltage
characteristics for currents applied perpendicular and parallel to the
columnar defects, at $T=T_{\rm BG}$.  In the limit of small currents
we observe Ohmic response when the nonlinear response length scale
exceeds the system size.  In a finite range of sufficiently small
currents the I-V characteristics show power-law behavior as indicated
by the straight lines in the plot.  The solid straight lines are given
by the scaling forms in Eqs.~(\ref{I-V-perp}) and (\ref{I-V-para}),
using the value $z=4.6$ determined in Fig.~\ref{rho} from the linear
resistivity.  For higher currents there are clearly visible
deviations, and a crossover to different power laws takes place.  If
this range of large currents is fitted to
Eqs.~(\ref{I-V-perp},\ref{I-V-para}) one has to assume different
values of $z$ in the $\perp$ and $\parallel$ directions: $z_{\perp}
\approx 6, z_{\parallel} \approx 4$, in agreement with
Ref.~\cite{wallin-girvin}.

\begin{figure}[tb]
\centerline{
\epsfxsize 0.5\columnwidth \epsfbox{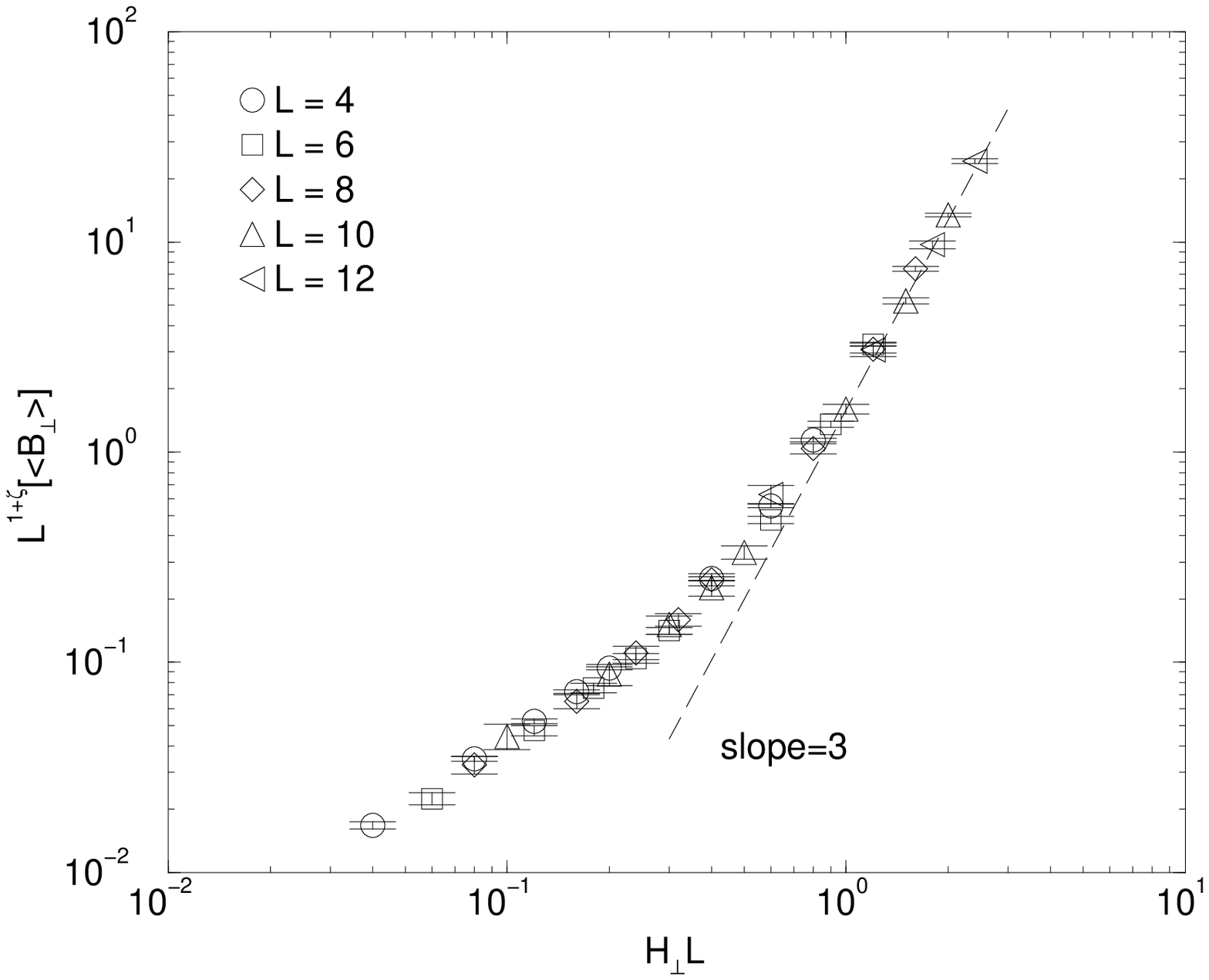}
\epsfxsize 0.5\columnwidth \epsfbox{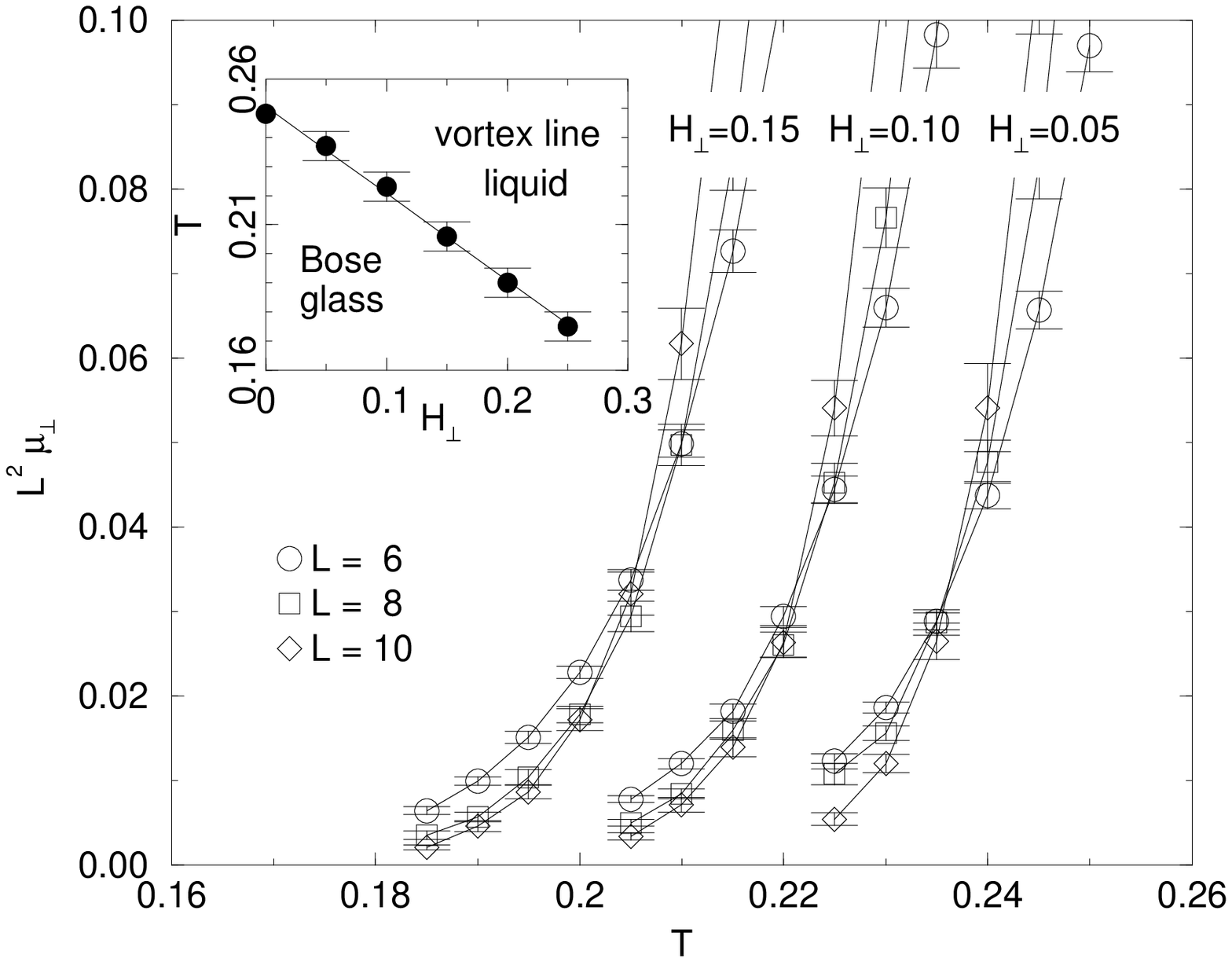}}
\caption{Finite size scaling of MC data for $B_\perp$ for various
sizes $L$ and applied perpendicular fields $H_{\perp}$ at $T=T_{\rm
BG}$.  The dashed line indicates the form $B_\perp \sim H_{\perp}^{1 +
\zeta }$, with $\zeta=2$, for $H_{\perp} L > 1$.  }
\label{B-H}
\end{figure}

\begin{figure}[tb]
\caption{MC data for magnetic permeability $\mu_\perp$ in a tilted
magnetic field.  The critical temperature $T_c(H_\perp)$ is roughly
estimated as the point where the curves for different $L$ intersect
(filled circles in the inset).  Inset: The resulting BG phase
boundary.  The solid straight line corresponds to a power law $T_{\rm
BG}-T \sim |H_{\perp}|^{1/\nu}$ with $\nu=1.0$.}
\label{scal_perm}
\end{figure}

We next consider tilted magnetic fields.  As before a constant field
was applied parallel to the columns, corresponding to a half filling,
and in addition we apply a small perpendicular field $H_{\perp}$.  We
first consider the effect of $H_\perp$ on $B_\perp$ at the Bose glass
temperature $T=T_{\rm BG}$, shown in Fig.\ \ref{B-H}.  The data
collapse for different system sizes and tilts verifies the scaling
form $B_\perp = L^{-1-\zeta}\tilde{B}_\perp(H_\perp L^{d-2})$.  The
dashed line corresponds to the power law scaling form $B_{\perp} \sim
H_{\perp}^{1+\zeta}$ for scaling variable $H_{\perp}L \gtrsim 1$.  To
locate the BG phase boundary as the magnetic field is tilted, we use
\Eq{mu-perp}, which, for finite systems and $d=3$, gives $\mu_\perp =
L^{-\zeta} \tilde{\mu}_{\perp} ([T-T_c(H_{\perp})] L^{1/\nu''})$.
Here $\nu''$ is the correlation length exponent for finite tilt
fields,
which belongs to a separate universality class~\cite{nelson-vinokur}.
At the critical temperature in a finite system, this should scale as
$\mu_\perp \sim L^{-\zeta}$ for large enough $L$.  In Fig.\
\ref{scal_perm} we use the crossing point for MC data for small system
sizes as a crude estimate of the critical temperature.  However,
rather large corrections to scaling are expected due to the closeness
of the BG fixed point,
and we were therefore not able to determine $\nu''$.  
The inset in Fig.\ \ref{scal_perm} shows the
transition temperatures (filled circles) vs.\ $H_{\perp}$.  The solid
straight line represents Eq.\ (\ref{line}), using the value $\nu=1.0$
for the BG transition \cite{dirtybosons}.  Thus the form of the BG
phase boundary given by Eq.\ (\ref{line}) seems consistent with the
simulation.

Finally we will compare our results with experiments.  Transport
measurements for Tl$_2$Ba$_2$CaCu$_2$O$_8$ (Tl-2212) thin films with
columnar defects and zero tilt \cite{phuoc} obtained $z \approx 4.9,
\nu \approx 1.1, \zeta \approx 1.9$, which closely agree with our
exponents.  Experiments for Tl-2212 with tilted magnetic fields
\cite{budhani} obtained $z'' \approx 4.4, \nu'' \approx 1.8$, where
the double prime indicates exponents for the tilted fixed point.
A heavily twinned YBCO single crystal, without artificial columnar
defects, gave \cite{grigera} $\nu(z-2)\approx 2.8$, for zero tilt,
which is close to our number $\nu(z-2)\approx 2.6$.  Quite different
experimental results were found in Ref.\ \cite{jiang}.  Experiments on
YBa$_2$Cu$_3$O$_7$ (YBCO) single crystals with columnar defects in low
magnetic fields (0-6.3 kOe) \cite{jiang} suggest $z \approx 2.2, \nu
\approx 1, \zeta \approx 1$.  The isotropic scaling indicates that the
vortex interaction is not effectively screened \cite{dirtybosons}, and
a model with a longrange vortex interaction may be suitable.  
We have also done some (limited) simulations for a model with
unscreened longrange interactions, and find a dynamical exponent $z
\approx 2$, which is considerably smaller than the one for strong
screening.
Experiments with tilted magnetic fields \cite{grigera,reed,jiang} show
a sharp cusp in the BG phase boundary, in agreement with the Bose
glass theory.  We note, however, that the actual shape of the BG phase
boundary in the figures in Refs.\ \cite{grigera,reed,jiang}, appears
to reasonably well fit our $T_{\rm BG}(0)-T_{\rm BG}(H_\perp) ~ \sim
|H_\perp|^{1/\nu}$ for small $H_\perp$, with exponent $1/\nu \approx
1.0$ instead of $1/3\nu$ from Ref.\ \cite{nelson-vinokur}.  More
experimental data for the precise shape of the BG phase boundary and
for other quantities would be useful to test our new scaling
relations.

In summary, we have analyzed the Bose glass transition in
superconductors with columnar defects by means of scaling theory and
Monte Carlo simulations.  For magnetic fields tilted away from the
direction of the columns we suggest a form of the BG phase boundary
for small tilts, which appears consistent with our simulations and
with experiments.  Our value $z = 4.6 \pm 0.3$ for the dynamical
critical exponent at the BG transition in a strongly screened model is
in good agreement with recent experiments.

\stars

We gratefully acknowledge very useful discussions with D.\ R.\ Nelson,
S.\ M.\ Girvin, and S.\ Teitel.  This work was supported by the Swedish
Natural Science Research Council, by the Swedish Foundation for
Strategic Research (SSF), and by Parallelldatorcentrum (PDC), Royal
Institute of Technology.

\end{document}

%% file: euromacr.tex
\def\And{{\rm and\ }}

\def\stars{\bigskip\centerline{***}\medskip}

\newif\ifboo \boofalse

\def\Review#1{\boofalse{\it #1},}
\def\Name#1{{\sc #1},}
\def\Vol#1{\ifboo Vol. {\bf #1}\else{\bf #1}\fi}
\def\Year#1{\ifboo #1\else(#1)\fi}
\def\Book#1{\bootrue{\it #1},}
\def\Page#1{\ifboo {\rm p. #1}\else{\rm #1}\fi}

%% file: column.bbl
\begin{thebibliography}{99}
\bibitem{foot} For $d \ne 3$ the expressions are not supposed to
agree.  Changing the dimension in the two problems means different
things, since topological defects in the phase of the superconducting
order parameter are lines only in 3D.
\end{thebibliography}
